\documentstyle[twocolumn,eqsecnum,aps,tighten,epsf]{revtex}

\newcommand{\expl}{\langle \!\langle}
\newcommand{\expr}{\rangle \!\rangle}

\begin{document}

\preprint{UGI-98-27}

\title{Stochastic interpretation of Kadanoff-Baym equations }

\author{Carsten Greiner\thanks{Talk presented at the 5th International Workshop
on Thermal Field Theories and their Applications, Regensburg, Germany,
August 1998}
and Stefan Leupold}

\address{Institut f\"ur Theoretische Physik, Justus-Liebig-Universit\"at
Giessen,\\
D-35392 Giessen, Germany}

\date{September 2, 1998}

\maketitle

\begin{abstract}
We show that the nonperturbative quantum
transport equations, the so called `Kadanoff-Baym equations',
within the non-equilibrium real time Green's function description
can be be understood as the ensemble average over stochastic equations of
Langevin type.
For this we couple a free scalar boson quantum field to
an environmental heat bath with some given temperature $T$.
The inherent presence
of noise and dissipation related by the fluctuation-dissipation theorem
guarantees that the modes or particles become thermally populated on average
in the long-time limit.
This interpretation leads to a more intuitive physical picture of the
process of thermalization and of the interpretation of the Kadanoff-Baym
equations.
One also immediately understands that the emerging wave equations
for long wavelength modes with momenta $|k|\ll T$
behave nearly as classical fields.
We also demonstrate how the problem of so called
pinch singularities is resolved by a clear physical necessity of damping
within the one-particle propagator. The occurrence of such
ill-defined terms arising solely in a strictly perturbative expansion
in out of equilibrium quantum field theory has a natural interpretation in analogy
to Fermi's golden rule.
\end{abstract}

\section{Motivation} \label{sec:intro}

Non-equilibrium many body theory had been traditionally
a major topic of research for describing various transport processes
in condensed matter physics and nuclear physics. Over the last years
a lot of interest for non-equilibrium quantum field theory has now emerged
in particle physics.
A very powerful diagrammatic tool is given
by the `Schwinger-Keldysh' \cite{Sc61} or `closed time path' (CTP)
technique by means of non-equilibrium Green's functions for describing a quantum system
also beyond thermal equilibrium \cite{Ch85}.
The resulting causal and nonperturbative equations of motion
(by various approximations),
the so called {\em Kadanoff-Baym equations} \cite{KB}, have to be considered
as an ensemble average over the initial density matrix
characterizing the preparation of the initial state of the system.
Typically, if the system is close to thermal equilibrium, the
(initial or resulting) density matrix allows for thermal fluctuations
which should be inherent to the transport process under consideration.
If fluctuations are physically present in the course of the evolution, then,
according to the famous fluctuation-dissipation theorem \cite{CW51},
also dissipation must be present. Or, in turn, if the (quantum) system behaves
dissipatively, as a consequence, there must exist fluctuations.
Again, the Kadanoff-Baym equations have to be understood
as an ensemble average over all the possible fluctuations.
This inherent stochastic aspect of the Kadanoff-Baym equations is what
we want to point out and thus provide, as we believe, some new physical insight
into its merely complex structure \cite{GL98a}.
Some of our conclusions are already known and can be found
to some extent in various different studies \cite{Ch85,DuB67}.
A new aspect, however, is the intimate connection of the Kadanoff-Baym equations
to Langevin like processes.

As a reminder of a Langevin process let us first briefly review
the description of classical Brownian motion. Consider a classical system described
by the variables $Q$ and $P$ interacting with a (heat) bath of oscillators
of various frequencies \cite{Co85} according to the hamiltonian
$$
H =  H_S(Q,P)  +  \sum_{\nu } (\frac{p_{\nu }^2}{2m_{\nu }}
      + \frac{m_{\nu }\omega _{\nu }^2}{2} q_{\nu }^2 )  - 
      Q \cdot \sum_{\nu } \Gamma _{\nu } q_{\nu }  \, .
$$
From the resulting equations of motion
the bath degrees of freedom can formally be integrated as
\begin{eqnarray}
q_{\nu }(t) & = & q_{\nu }(0) \cos \omega _{\nu } t  + 
\frac{p_{\nu } (0)}{m_{\nu } \omega _{\nu }} \sin \omega _{\nu } t  
\nonumber \\ && {}  + 
\int_0^t dt' \, \frac{ \sin \omega _{\nu } (t-t')}{m_{\nu } \omega _{\nu }}
\Gamma _{\nu } \, Q(t')    \, ,
\label{bathdof}
\end{eqnarray}
so that the systems variables obey the effective
Langevin-like equation of motion
\begin{equation}
\label{systemdof}
\ddot{Q}  +  \frac{\partial H_S^{(m)}}{\partial Q}  + 
2 \int_0^t dt' \, \Gamma (t-t') \dot{Q}(t')  =  \xi (t)  \, .
\end{equation}
The bare hamiltonian $H_S$ is slightly modified as
\begin{equation}
H_S^{(m)} (Q,P)   =   H_S (Q,P)  + 
\left( \sum _{\nu } \frac{(-\Gamma _{\nu } ^2)}{ 2 m_{\nu } \omega _{\nu }^2} 
\right)
Q^2  \, .
\end{equation}
On the other hand a Stokes-type frictional term
(in general nonlocal in time) with a dissipation kernel
\begin{equation}
\label{dissipkernel}
\Gamma (t-t')   =  \frac{1}{2}
\sum _{\nu } \frac{\Gamma _{\nu } ^2}{ m_{\nu } \omega _{\nu }^2} 
\cos \omega _{\nu } (t-t')
\end{equation}
enters into the left hand side of (\ref{systemdof}) supplemented by
a source term $\xi (t)$ on the right hand side being given as
\begin{eqnarray}
\xi (t)  & = &
\sum _{\nu } \Gamma _{\nu } \left( [q_{\nu }(0) - \frac{\Gamma _{\nu }}{ m_{\nu } 
\omega _{\nu }^2}
Q(0) ] \cos \omega _{\nu } t 
\right. \nonumber \\ && \left. \phantom{mmmmmmmmmmm}
+ \frac{p_{\nu }(0)}{ m_{\nu } \omega _{\nu }} \sin \omega _{\nu } t \right)  \, .
\label{noiseclassic}
\end{eqnarray}
In $\xi $ especially the initial conditions of the bath oscillators
enter. In general, they are not known exactly, but will follow
some statistic distribution. Hence, $\xi (t)$ has to be interpreted as a stochastic
or `noisy' source driving the fluctuations of the systems degrees of freedom.
Indeed, if the initial conditions are randomly choosen by a thermal distribution
$ P[q(0),p(0)]  \sim e^{-H_B/k_BT} $ \cite{Co85}, $\xi (t)$ is completely
specified by a Gaussian distribution
(`central limit theorem')
with zero mean and the correlation kernel $I$
\begin{eqnarray}
\expl \xi(t) \expr & = & 0  \, , \nonumber \\
I(t-t')  :=  \expl \xi(t) \xi(t')\expr & = & k_B T
\sum _{\nu } \frac{\Gamma _{\nu } ^2}{ m_{\nu } \omega _{\nu }^2} 
\cos \omega _{\nu } (t-t')  
\nonumber \\ 
& \equiv &  2 k_B T \Gamma (t-t')   \, .
\label{BMnoise1}
\end{eqnarray}
$\expl \ldots \expr$ denotes the average over all possible realizations of the
stochastic variable $\xi (t)$.
The dissipation kernel $\Gamma $ and the random force $\xi (t)$
have the {\em same} microscopic origin. In fact, as just derived,
there exists a simple relation between them, called the
fluctuation-dissipation theorem \cite{CW51}. The evolution of the systems
degrees of freedom are thus supplemented by an interplay between a
dissipative term and a fluctuating source. On the average, they should then
equilibrate thermally. As a special application consider
a heavy `Brownian' particle with mass $M$ placed in a thermal environment
obeying an effective Langevin equation as just stated, i.e.
\begin{equation}
M \ddot{x}  + 
2\int\limits_{-\infty}^{t} dt' \, \Gamma (t-t') \, \dot{x}(t') 
=  \xi (t)    \, .
\label{Lang2}
\end{equation}
(Here we assume that the friction kernel $\Gamma (t)=\Gamma (-t)$
is `well behaved'
meaning that its Fourier transform $\bar\Gamma (\omega ) \geq 0$
and has only some finite extension in time.)
To convince oneself that such a description
in the long time limit is in accordance with the
equipartition condition
$\expl p^2 \expr/(2M) \stackrel{!}{=} T/2$,
one has to solve (\ref{Lang2}) by means of the retarded and advanced
Green's function:
\begin{eqnarray}
\bar G_{\rm ret} (\omega )   & = &
\frac{1}{-i\omega + \left( \frac{1}{M} \bar \Gamma (\omega )  + \frac{i}{\pi M}
{\it P} \int d\omega '\frac{\bar \Gamma (\omega') }{\omega - \omega'} + \epsilon
\right) }  
\nonumber \\ & = &
\bar G_{\rm av}^{*} (\omega )   \, .
\label{Gret}
\end{eqnarray}
One now finds in the long-time limit on the average the
desired property
\begin{eqnarray}
\lefteqn{\expl p^2(t\rightarrow \infty ) \expr}
\nonumber \\  & = &
\int\limits _{- \infty }^{\infty } dt_1 dt_2 \,
G_{\rm ret}(t-t_1) \expl \xi (t_1) \xi (t_2) \expr G_{\rm av} (t_2-t)
\nonumber \\
& = & \int \frac{d\omega }{2\pi } \, \bar G_{\rm ret}(\omega ) \bar I (\omega ) 
\bar G_{\rm av} (\omega)
 =  T \, M   \, .
\label{equi1}
\end{eqnarray}
Hence, irrespective of the detailed form of the
friction kernel $\Gamma (t)$, the equipartition condition is automatically
fulfilled as long as the noise kernel $I(t)$ fulfills the
fluctuation-dissipation relation (\ref{BMnoise1}).

For a further physical motivation let us now return
to quantum field theory and already point out some similarities.
One of the major present topics in quantum field theory at finite temperature
or near thermal equilibrium concerns the evolution and behavior of the long
wavelength modes. These modes often lie entirely in the non-perturbative regime.
Therefore solutions of the classical field equations in Minkowski space have
been widely used in recent years to describe long-distance properties
of quantum fields that require a non-perturbative analysis.
A justification of the classical treatment of the long-distance dynamics
of bosonic quantum
fields at high temperature is based on the observation that the average 
thermal amplitude of low-momentum modes is large.  For a weakly coupled 
quantum field the occupation number of a mode with wave vector $\vec{ p}$
and frequency $\omega_{\vec{ p}}$ is given by the Bose distribution
\begin{equation}
n(\omega_{\vec{ p}}) = \left(e^{\omega_{\vec{ p}}/T} -1 \right)^{-1}.
\label{Bose}
\end{equation}
For temperatures $T$ much higher than the (dynamical) mass scale $m^*$
of the quantum field, the occupation number
for the low-momentum modes $|\vec{p }|\ll T$
becomes large and
approaches the classical equipartition limit
\begin{equation}
n(\omega_{\vec{ p}}) \buildrel \vert {\bf p}\vert \to 0 \over \longrightarrow
{T\over m^*} \gg 1. \label{climit}
\end{equation}
The classical field equations
(which can be understood as a coherent state approximation to the full quantum 
field theory)
should provide a good approximation for
the dynamics of such highly occupied modes.
At a closer look, however, the cogency of this heuristic argument
suffers considerably.  The thermodynamics of a classical 
field is only defined if
an ultraviolet cut-off $k_c$ is imposed on the momentum {\bf p} such as a
finite lattice spacing $a$.
Many, if not most, thermodynamical
properties of the classical field depend strongly on the value of the 
cut-off parameter $k_c$ and diverge in the continuum limit $(k_c \to\infty)$.
In a correct semi-classical treatment of the soft modes the hard modes
thus cannot be neglected, but it
should incorporate their influence in a consistent way.
In a recent paper of one of us \cite{CG97} it was shown how to construct
an effective semi-classical action for describing not only the classical
behavior of the long wavelength modes below some given
cutoff $k_c$,
but taking into account also perturbatively the interaction among the soft
and hard modes. By integrating out the `influence' of the hard modes
on the two-loop level for standard $\phi^4$-theory the emerging semi-classical
equations of motion
for the soft fields
can be derived from an effective action
and become stochastic equations of motion
of Langevin type \cite{CG97}:
\begin{eqnarray}
\lefteqn{{\partial^2\phi\over\partial t^2} +  \left ( {\bf k}^2+ \tilde m^2 
+ \!\!\! \sum_{i=a,b,c} \mu_{1,k_c}^i \right)\phi + \left(
{\tilde g^2\over 6}+\mu_{2,k_c}^{(d)} \right) \otimes \phi^3} 
\nonumber \\ 
&& {} + \mu_{3,k_c}^{(e)} \otimes \phi^5 +\sum_{i=c,d,e} \eta^{(i)}_{k_c}
\dot{\phi } \;
\approx \; \sum_{N=1}^3 \xi_n \otimes \phi^{N-1}\;. \label{Lang3}
\end{eqnarray}
(Here $\otimes$ denotes a convolution in momentum space.  The coefficients
$\mu_i$ as well as the damping coefficients $\eta_i$
depend on {\bf k} and $t$, as well as on the momentum cut-off $k_c$.)
These resemble in its structure the analogous expression to eq.~(\ref{Lang2}).
The hard modes act as an
environmental heat bath. They also guarantee that the soft modes become, on average,
thermally populated with the same temperature as the heat bath.
Equivalently to (\ref{equi1}) we will see that a similar relation exists
for the average fluctuation of the amplitude squared.
For the semi-classical regime where $|\vec{ p}|\ll T$ it
will yield
\begin{eqnarray}
\expl |\phi (\vec{ p},t\rightarrow \infty )|^2 \expr &  \approx &
\frac{V}{E_p^2} \, T  
\approx 
\frac{V}{E_p} \, n(\vec{p},t)  
\, ,
\label{equi2}
\end{eqnarray}
where $V$ is the volume of the system.
Such kind of Langevin description for
the non-perturbative evolution of (super-)soft modes
(on a scale of $|\vec{p}| \sim g^2 T$)
in non Abelian gauge theories
has recently been put forward \cite{HS97}.

In analogy to the Langevin description stated above
we want to formulate in the following the effect of the heat bath on
the evolution of the system degrees of freedom by means of the CTPGF technique.
It is the intention of our study to provide in a self-contained way
new insight into the dissipative and
stochastic character of the CTPGF approach in thermal and non-equilibrium
field theory.
For this we discuss a free scalar field theory
interacting with a heat bath. In the next section
we first review the CTPGF technique and evaluate the average
characteristic properties
of the CTP propagator resulting from the (non-equilibrium) equations of motion,
the Kadanoff-Baym equations.
Within a rather simple rearrangement of the
interaction kernels stemming from the heat bath,
it becomes transparent to identify
the dynamically generated mass shift, the dissipation
and the fluctuation terms, all of them contributing to the dynamical
evolution of the system.
We also show that the fluctuation-dissipation theorem emerges
naturally and is thus stated in microscopic terms.
We then introduce in the third section the concept of
a stochastic generating functional with external noise.
In this way the analogy of thermalization in quantum field theory
to a Langevin-like process becomes apparent.
We thus can point out the influence of the
noise correlation function on the propagator which contains the
occupation number. As motivated above, one also immediately realizes
that the long wavelength modes with momenta $|\vec{k}| \ll T$
and energies $|\omega |\ll T$ behave
as classical propagating modes for a weakly interacting theory.
We also briefly mention an attempt how to
to derive an effective Boltzmann-Langevin equation 
for the phase-space occupation number of the bosonic particles 
from first principles.
Section IV is devoted to the issue of so called pinch singularities
emerging in a strictly perturbative evaluation in out of equilibrium
quantum field theory. We will give first a very clear physical
picture for their occurence. We then show
how the 'problem' of pinch singularities is cured
naturally within the (non-perturbative) processes of
thermalization and damping and providing a bridge to standard kinetic theory.
We will end our findings with a brief summary and conclusions in section V.

\section{Kadanoff-Baym equations } \label{sec:CTPGF}

In this section we study the thermalization of a simple quantum field theory.
For this purpose we couple a system of free scalar fields to an
environmental heat bath of temperature $T$.
We start with the closed time path action for this scalar field $\phi$ \cite{GL98a}:
\begin{eqnarray} 
S &=& 
\int \!\!d^4\!x
{1\over 2} \left[ 
\phi^+ \, (-\Box-m^2) \,\phi^+ - \phi^- \, (-\Box-m^2) \,\phi^- 
\right. \nonumber \\ &&  \phantom{mmmm}
{}- \phi^+ \,\Sigma^{++} \,\phi^+ - \phi^+ \,\Sigma^{+-} \,\phi^- 
\nonumber \\ && \phantom{mmmm} \left.
- \phi^- \,\Sigma^{-+} \,\phi^+ - \phi^- \,\Sigma^{--} \,\phi^-
\right] 
\,.
\label{eq:ctpac}
\end{eqnarray}
The time integration starts at some fixed time $t_0$ at which
the system starts to evolve from some initial density matrix.
The interaction among the system and the heat bath is stated by
an interaction kernel involving a self energy (or `mass') operator
$\Sigma $ resulting effectively from integrating out the heat bath degrees
of freedom.
Clearly, this self energy operator is the only quantity which might drive
the system towards equilibrium.
We assume that $\Sigma $ is solely determined by the properties
of the heat bath. Ideally such a scenario holds for the linear response regime.

In (\ref{eq:ctpac}) the self energy contribution from the heat bath
is parametrized in the Keldysh notation by the four self energy parts
\begin{eqnarray}
\lefteqn{\Sigma^{++}(x_1,x_2)  = } \nonumber \\  
  \label{eq:selfpp}
&& \Theta(t_1-t_2)\,\Sigma^>(x_1,x_2) + \Theta(t_2-t_1)\,\Sigma^<(x_1,x_2)
\\
\lefteqn{\Sigma^{--}(x_1,x_2)  = } \nonumber \\  
&& \Theta(t_2-t_1)\,\Sigma^>(x_1,x_2) + \Theta(t_1-t_2)\,\Sigma^<(x_1,x_2)
\\
\lefteqn{\Sigma^{+-}(x_1,x_2)  =  -\Sigma^<(x_1,x_2)}
\\
\lefteqn{\Sigma^{-+}(x_1,x_2)  =  -\Sigma^>(x_1,x_2)}
  \label{eq:selfmp}
\end{eqnarray}
where $t_1$ and $t_2$ are the time components of the vectors $x_1$ and $x_2$,
respectively. With these definitions at hand one sees that in principle
only two self energy parts are independent, e.g. $\Sigma^<$ and $\Sigma^>$.
One can also easily show that
the difference $\Sigma^>(x_1,x_2) - \Sigma^<(x_1,x_2)$ is real while the sum
$\Sigma^>(x_1,x_2) + \Sigma^<(x_1,x_2)$ is purely imaginary. 

If, as assumed, the heat bath stays at thermal equilibrium
at some fixed temperature $T$ and the self energy is solely
determined by the property of this heat bath, then the
important relation
\begin{equation}
  \bar \Sigma^>(k) = e^{k_0/T}\, \bar \Sigma^<(k) \,,  \label{eq:heat}
\end{equation}
holds. This property (\ref{eq:heat}) which is nothing but the Kubo-Martin-Schwinger
boundary condition \cite{KMS,Ch85,La87} will become crucial below when we
will show that the modes of our system in the long-time limit 
will indeed equilibrate at the  temperature of the heat bath.

In order to explicitly explore the causal structure, it is useful
to introduce the more physical retarded and advanced self energies
\begin{eqnarray}
\Sigma^{\rm ret}(x_1,x_2) & := & 
\Theta(t_1-t_2)\,\left[ 
\Sigma^>(x_1,x_2) - \Sigma^<(x_1,x_2) 
\right]   \,, \nonumber \\
\label{eq:defsigav}
\Sigma^{\rm av}(x_1,x_2) & := & 
\Theta(t_2-t_1)\,\left[ 
\Sigma^<(x_1,x_2) - \Sigma^>(x_1,x_2) 
\right]  \,. \nonumber 
\end{eqnarray}

Given the fields $\phi^+$ and $\phi^-$ on the two branches of the CTP contour,
four two-point functions can be defined:
\begin{equation} 
\label{eq:proptpf} 
D^{ab}(x_1,x_2) := -i \langle \phi^a(x_1) \,\phi^b(x_2) \rangle \,, \quad 
a,b = +,- \,.
\end{equation}
The expectation value is defined via path integrals as
\begin{equation}
  \label{eq:expec1}
  \langle {\cal O} \rangle := {1\over N} \int\!\! {\cal D}[\phi^+,\phi^-] \, 
{\cal O} \, e^{iS[\phi^+,\phi^-]} \rho[\phi^+,\phi^-] 
\end{equation}
where $\rho$ is the density matrix of the system for the initial time $t_0$.

When evaluating the two-point functions by using (\ref{eq:expec1})
one finds
\begin{eqnarray}
\lefteqn{D^{++}(x_1,x_2)  = } \nonumber \\ &&
\Theta(t_1-t_2)\,D^>(x_1,x_2) + \Theta(t_2-t_1)\,D^<(x_1,x_2) \,, \\
\lefteqn{D^{--}(x_1,x_2)  = } \nonumber \\ &&
\Theta(t_2-t_1)\,D^>(x_1,x_2) + \Theta(t_1-t_2)\,D^<(x_1,x_2)
\,, \\
\label{eq:propagators3}
\lefteqn{D^{+-}(x_1,x_2)  =  -i \langle \phi^-(x_2) \,\phi^+(x_1) \rangle
 =:  D^<(x_1,x_2) \,, } \\
\label{eq:propagators4}
\lefteqn{D^{-+}(x_1,x_2)  =  -i \langle \phi^-(x_1) \,\phi^+(x_2) \rangle
 =:  D^>(x_1,x_2) \,. }
\end{eqnarray}
In addition, one also now introduces retarded and advanced quantities by
\begin{eqnarray}
D^{\rm ret}(x_1,x_2) & = &
\Theta(t_1-t_2)\,\left[ 
D^>(x_1,x_2) - D^<(x_1,x_2) \right]  \, ,
\nonumber \\
D^{\rm av} (x_1,x_2)& = &
\Theta(t_2-t_1)\,\left[ 
D^<(x_1,x_2) - D^>(x_1,x_2) \right]  \, .
\nonumber 
\end{eqnarray}
From (\ref{eq:ctpac}) it is straightforward
to obtain the equations of motion for 
the matrix elements.
Since the four two-point functions 
are not independent of each other the four equations will contain redundant
information. Therefore we will state the equations of motion for retarded and
advanced propagator and for an off-diagonal element, say 
\begin{equation}
  \label{eq:defpropkl}
  D^< := D^{+-} \,. 
\end{equation}
The equation of motion for the retarded propagator then reads
\cite{DuB67,Ch85,GL98a}
\begin{equation}
  \label{eq:eomretprop}
 (-\Box -m^2-\Sigma^{\rm ret}) D^{\rm ret} = \delta
\end{equation}
A similar one is given for the advanced propagator.
Please note already the analogy to the solving of the Langevin equation
carried out in the introduction by introducing the retarded and advanced
propagator.
Additional information now comes from the equation of motion of
the propagator $D^<$, which reads
\begin{equation}
  \label{eq:kadbaym}
(-\Box -m^2)  D^< - \Sigma^{\rm ret} D^< - \Sigma^< D^{\rm av} = 0 \,, 
\end{equation}
This is just the famous {\em Kadanoff-Baym} equation \cite{KB}.

(\ref{eq:eomretprop}) and (\ref{eq:kadbaym}) determine the complete and causal
(non-equilibrium) evolution for the two-point functions.
To get now more physical insight into the (effective) action
given in (\ref{eq:ctpac})
and in the resulting Kadanoff-Baym equations we
now introduce the following {\em real} valued quantities:
\begin{eqnarray}
  s(x_1,x_2) & := & {1\over 2} {\rm sgn}(t_1-t_2) 
\left(\Sigma^>(x_1,x_2) - \Sigma^<(x_1,x_2) \right) 
\nonumber \\ 
\label{eq:sdef}
& = & s(x_2,x_1) \, ,  \\ 
a(x_1,x_2) & := & {1\over 2} \left(\Sigma^>(x_1,x_2) - \Sigma^<(x_1,x_2) \right) 
\nonumber \\
\label{eq:adef}
& = & - a(x_2,x_1) \, , \\ 
I(x_1,x_2) & := & -{1\over 2i} \left(\Sigma^>(x_1,x_2) + \Sigma^<(x_1,x_2) 
\right)  
\nonumber \\
\label{eq:Idef}
& = &  I(x_2,x_1)
\,.
\end{eqnarray}
Our notion for $s$ and $a$ serves as a reminder for the respective symmetry 
properties. It basically represents the standard decomposition of the real
and imaginary part of the Fourier transform of the retarded self energy
operator $  \bar \Sigma^{\rm ret}$ \cite{DuB67,Ch85}.
$s$ yields a (dynamical) {\em mass shift} for the $\phi$
modes caused by the
interaction with the modes of the heat bath, while $a$ is responsible for the
damping, i.e. {\em dissipation}
of the $\phi$ fields. The important thing to point out will be that
$I$ characterizes the {\em fluctuations}.

We first note that on account of (\ref{eq:sdef})-(\ref{eq:Idef}) together with
(\ref{eq:selfpp})-(\ref{eq:selfmp}) the CTP action (\ref{eq:ctpac})
can be written as
\begin{eqnarray}
  S & = & {1\over 2} \left[ 
\phi^+ \, (-\Box-m^2) \,\phi^+ - \phi^- \, (-\Box-m^2) \,\phi^- 
\right. \nonumber \\   
&&  \phantom{mm}
{}- (\phi^+ - \phi^-)\, (s+a) \, (\phi^+ + \phi^-) \nonumber \\
\label{eq:IFac}   
&& \left. \phantom{mm}
+ i \, (\phi^+ - \phi^-) \, I \, (\phi^+ - \phi^-) 
\right]  \,.
\end{eqnarray}
This expression is
identical to the {\em influence functional} given by Feynman and
Vernon \cite{Fe63,S82}.
To the exponential factor in the path integral 
(\ref{eq:expec1}) the $(s+a)$ term contributes a phase while the $I$ term 
causes an exponential damping and thus signals nonunitary evolution.
Following the ideas of Feynman and Vernon one can therefore indeed 
identify $I$ as a `noise'
correlator (see next section).
The equivalence of the CTP formalism and the influence functional approach
has indeed been pointed out already in \cite{SCYC88} on a strictly formal level.
Our discussion here is inspired on more physical intuition.

The two relevant equations of motion are stated as
\begin{eqnarray}
  \label{eq:eomretsa}
(-\Box -m^2-s-a) D^{\rm ret}  & = & \delta \,, \\
  \label{eq:eomklsaI}
(-\Box -m^2-s-a)  D^< + (a+iI) D^{\rm av} & = & 0 \,.
\end{eqnarray}
We see that the last equation (the `Kadanoff-Baym' equation)
is the only one where $I$ occurs. Hence the
retarded (and advanced) propagators are determined by $s$ and $a$ while the
number density is additionally influenced by the fluctuations $I$.

For the interpretation of $s$ and $a$
we consider the long-time behavior of these equations.
In this case we can assume that the system becomes translational
invariant in time and space and the boundary terms are no longer important. 
For $\bar D^{\rm ret}$ one immediately finds
\begin{equation}
  \label{eq:dretimp}
\bar D^{\rm ret}(k) = {1 \over k^2-m^2-\bar s(k)-\bar a(k)} =
\left( \bar D^{\rm av}(k) \right) ^{*} \,.
\end{equation}
From this the spectral function is given by
\begin{eqnarray}
{\cal A}(k) & := &
{i\over 2} [\bar D^{\rm ret}(k) - \bar D^{\rm av}(k)] 
\nonumber \\
& = & {i\, \bar a(k) \over [k^2-m^2-\bar s(k)]^2 + \vert \bar a(k) \vert ^2 } \,. 
  \label{eq:defspec}
\end{eqnarray}
Note that $\bar a(k)$ is purely imaginary since $a(x_1,x_2)$ is real and 
antisymmetric. For the same reason $\bar s$ is real.
Inspecting the spectral function it becomes obvious that
$\bar s\equiv {\rm Re } \bar \Sigma ^{\rm ret}$
contributes an 
(energy dependent) {\em mass shift} while
$\bar a \equiv i{\rm Im } \bar \Sigma ^{\rm ret}$
causes the {\em damping} of propagating
modes \cite{KB,DuB67}. $\bar a$ is related to the commonly used
damping rate $\bar{ \Gamma}$ via
\begin{equation}
  \label{defdamprate}
\bar{ \Gamma} (k)  =  i \frac{\bar{a}(k)}{k_0}   \, .
\end{equation}

For $D^<$ one finds in the long-time limit the relation
\begin{eqnarray}
\bar D^<(k) &= & \bar D^{\rm ret}(k)
\bar \Sigma^{<} \bar D^{\rm av}(k)  \nonumber  \\
&= & \bar D^{\rm ret}(k)
[-\bar a(k)-i\bar I(k)] \bar D^{\rm av}(k)   \nonumber \\
&=&
{-\bar a(k)-i\bar I(k) \over i\, \bar a(k)}\, {\cal A}(k) \,
\equiv   -2i \,n(k) {\cal A}(k)  \, .
  \label{eq:dkltherm}
\end{eqnarray}
We note that the first equation stated in (\ref{eq:dkltherm})
is sometimes denoted
in the literature \cite{DuB67,Ch85}
as a generalized fluctuation-dissipation theorem.
We will now outline its intimate connection
to a standard Langevin process in the following.

To proceed we need a 
relation between $\bar a$ and $\bar I$, i.e.~a relation between the damping and 
the noise term. In our case
it is a simple consequence of the KMS condition (\ref{eq:heat}) using the 
definitions (\ref{eq:adef}), (\ref{eq:Idef}): 
\begin{equation}
  \label{eq:finpnd}
n(k)  = 
{\bar \Sigma^<(k) \over \bar\Sigma^>(k) -\bar\Sigma^<(k) } 
= {1 \over e^{k_0/T} -1 }
 \equiv  n_B(k_0)  \, ,
\end{equation}
which indeed shows that the phase space occupation number of the soft modes in 
the 
long-time limit becomes a Bose distribution with the temperature
of the heat bath.
(\ref{eq:finpnd}) is independent
of the explicit and detailed form of the self energy but is solely determined
by the KMS condition.
If one now assumes that the coupling between the bath and the system
becomes very weak, i.e. $\bar{a},\bar{\Gamma }, \bar{I} \rightarrow 0$,
the expression (\ref{eq:finpnd})
\begin{equation}
  \label{eq:finpndweak}
n(k)  \left(\rightarrow \,
\frac{'0'}{'0'}\right) \, \rightarrow \,
{1 \over e^{k_0/T} -1 }
\end{equation}
is still preserved as long as the KMS condition is fulfilled.

It is now very illuminating to explicitly write down the relation between 
$\bar a(k)$ and $\bar I(k)$
\begin{eqnarray}
  \label{eq:fludis}
\bar I(k) = 
{\bar\Sigma^>(k) +\bar\Sigma^<(k) \over \bar\Sigma^>(k) -\bar\Sigma^<(k)} \, 
i \, \bar a(k) = {\rm coth}\!  \left({k_0 \over 2T}\right) i \, \bar a(k) \,. 
\end{eqnarray}
In the high temperature (classical) limit one gets
\begin{equation}
  \label{eq:hight}
  \bar I(k) = {T \over k_0} 2i\, \bar a(k) \,,
\end{equation}
or, employing (\ref{defdamprate}),
\begin{equation}
  \label{eq:hight1}
  \bar I(k) = 2T\, \bar{\Gamma }(k) \,.
\end{equation}
Recalling our discussion of Brownian motion in the introduction
this compares favorably well with (\ref{BMnoise1})!
Indeed as we shall see in the next section we can define a quantity which 
obeys a Langevin equation very similar to the one for the Brownian particle.
The physical meaning of $I$ as a `noise' correlator will become obvious.
The relation (\ref{eq:fludis}) thus already represents the
{\em generalized fluctuation-dissipation relation}
\cite{Co85} from a microscopic point of view
by the various definitions of $\bar{I},\,  \bar{a}$ and $\bar{\Gamma }$
through the parts $\bar{\Sigma }^<$ and $\bar{\Sigma }^>$ of the self energy.

We close this section by remarking that in all applications typically
the major goal of the Kadanoff-Baym equations had been to derive a standard
kinetic transport equation for the (semi-classical) phase-space distribution
$f(\vec{x}, \vec{k},t)$ which should be valid
for weak coupling and a nearly homogeneous system.
If the system is in a general off-equilibrium state the two-point functions
depend not only on the so called microscopic variable 
$u:=x_1$$-$$x_2$ but additionally on the macroscopic center-of-mass variable 
$X:=(x_1+x_2)/2$.
Performing a standard Wigner transformation for $D^<$ and making
a gradient expansion up to linear order in the center-of-mass variable $X$
one finds the well-known relativistic kinetic transport equation
in the quasi-particle approximation for the semi-classical
(onshell) one-particle phase-space distribution  \cite{DuB67,Ch85,GL98a}
\begin{eqnarray}
  \label{eq:boltz}
\lefteqn{k_\mu \partial^\mu_X f(\vec{x},\vec{k},t)
=}  \\ &&
 {1\over 2} \left. \left(
i\bar \Sigma^<(X,k) \, [f(\vec{x},\vec{k},t)+1] - i\bar\Sigma^>(X,k) \,
f(\vec{x},\vec{k},t)
\right) 
\, \right|_{k_0=\omega^0_{\vec{k}}}  \nonumber 
\end{eqnarray}
resembling in its form the Boltzmann equation.
Here all energies have to be evaluated onshell.
In this form it is obvious that $i\bar \Sigma^< / 2 \omega^0_{\vec{k}}$
can be interpreted as the production rate, while
$i\bar \Sigma^> / 2 \omega^0_{\vec{k}}$ stands for the absorbtion rate
for modes with the respective energy.

\section{Stochastic interpretation and some consequences}
\label{sec:langev}

To see the connection between the formalism presented in
the previous section \ref{sec:CTPGF}
and stochastic equations we decompose the influence action $S$
as given in (\ref{eq:IFac})
in its real and imaginary part and rewrite the generating functional
\cite{S82,GL98a}
\begin{eqnarray}
\lefteqn{Z[j^+,j^-]} \nonumber \\
& := & \int\!\! {\cal D}[\phi^+,\phi^-] \,
\rho[\phi^+,\phi^-]\, 
\nonumber \\ && \phantom{m} \times
e^{i{\rm Re}S[\phi^+,\phi^-] +i\,j^+\phi^+ + i\,j^-\phi^-
   -{1\over 2} (\phi^+ - \phi^-) \, I \, (\phi^+ - \phi^-) } 
\nonumber \\
& = & {1\over \tilde N} \int\!\! {\cal D}[\xi ] \,
       e^{-{1\over 2}\xi \,I^{-1}\xi } \,
      \int\!\! {\cal D}[\phi^+,\phi^-] \,
\rho[\phi^+,\phi^-]\, 
\nonumber \\ && \phantom{m} \times
e^{i{\rm Re}S[\phi^+,\phi^-] +i\,j^+\phi^+ + i\,j^-\phi^- 
   +i\,\xi\,(\phi^+ - \phi^-) }
\nonumber \\
& = & {1\over \tilde N} \int\!\! {\cal D}[\xi ] \,
       e^{-{1\over 2}\xi \,I^{-1}\xi } \,
       Z'[j^+ +\xi ,j^- -\xi ]  \,
\nonumber \\
  \label{eq:ZZpr}
& \equiv &  \expl Z'[j^+ +\xi,j^- -\xi]\, \expr
\end{eqnarray}
with 
\begin{equation}
  \label{eq:defnorm2}
\tilde N := \int\!\! {\cal D}\xi \, 
e^{-{1\over 2}\xi \,I^{-1}\xi } \,.
\end{equation}
The generating functional $Z[j^+,j^-]$ can thus be
interpreted as a new stochastic
generating functional $Z'[j^+ +\xi,j^- -\xi]$ averaged over a random 
(noise) field $\xi$ which is Gaussian distributed with the width function $I$, 
i.e. 
\begin{equation}
  \label{eq:defavdbl}
\expl {\cal O} \expr := 
{1\over \tilde N}
\int\!\! {\cal D}\xi \, {\cal O}\, e^{-{1\over 2}\xi \,I^{-1}\xi } \,. 
\end{equation}
From the last definition we find that the (ensemble) average
over the noise field vanishes
\begin{equation}
  \label{eq:noiseav}
\expl \xi \expr = 0  \, ,
\end{equation}
while the noise correlator is given by 
\begin{equation}
  \label{eq:noisecor}
\expl \xi \xi \expr = I \, .
\end{equation}
The action entering the definition of $Z'$ is no longer $S$, but only the real
part of the influence action (\ref{eq:IFac}).

From this new stochastic functional $Z'$
a {\em Langevin} equation for a classical $\phi$ field can be derived \cite{GL98a}.
Noting that the fields
$\langle\phi^+ \rangle_\xi$ on the upper branch and
$\langle\phi^- \rangle_\xi$ on the lower branch are equal
(and denoted as $\phi_\xi$ in the following),
its equation of motion derived from $Z'$ takes the form
\begin{equation}
  \label{eq:eomcl2}
(-\Box-m^2-s) \, \phi_\xi- a \, \phi_\xi= -\xi \,. 
\end{equation}
This, indeed, represents a standard Langevin equation.
The spatial Fourier transform of the Langevin equation (\ref{eq:eomcl2})
then takes the form
\begin{eqnarray}
\lefteqn{\ddot{\phi}_\xi (\vec{k},t)
+(m^2+\vec{k}^2-2\Gamma (\vec{k},\Delta t=0)) \phi_\xi (\vec{k},t) }
\nonumber \\ &&
+ 2\int\limits_{-\infty}^{t} dt' \, \Gamma (\vec{k},t-t') \,
 \dot{\phi}_\xi (\vec{k},t')
 =  \xi (\vec{k},t)  \, ,
  \label{eq:eomcl3}
\end{eqnarray}
The analogy between this Langevin equation (\ref{eq:eomcl3}) and the
one for a single classical oscillator is obvious.
The important difference, however, is the fact that the
corresponding relations (\ref{eq:fludis}) and (\ref{BMnoise1})
between the respective noise kernel $I$ and friction kernel $\Gamma $
only agree in the high temperature limit.

One can now ask to what extend the classical equations of motion
(\ref{eq:eomcl2}) together with (\ref{eq:noisecor})
are an approximation for the full quantum problem given by the
equation of motion (\ref{eq:eomklsaI}) for $D^<$.
Inverting (\ref{eq:eomcl2}) one finds for the correlation function
in the long-time limit
\begin{eqnarray}
  \label{eq:clapp}
-i\expl \, \langle \phi^+ \rangle_\xi \, \langle \phi^- \rangle_\xi \expr 
= -i \, D^{\rm ret} \expl \xi \xi \expr D^{\rm av} 
= -i \, D^{\rm ret} I\, D^{\rm av} \,.
\end{eqnarray}
Note that (\ref{eq:clapp}) is indeed the relation (\ref{equi2})
advocated in the introduction to hold in the
(semi-)classical regime.
This one has to compare with the full quantum correlation function
$D^<$ of (\ref{eq:dkltherm}).
One thus has that $(-\bar a-i\bar I)$ is approximated by $-i\bar I$.
Of course this is justified, if 
$\vert \bar a \vert \ll \bar I$
holds. Using the microscopic quantum version (\ref{eq:fludis}) of the 
fluctuation-dissipation theorem this is equivalent to
${\rm coth}\left( {k_0 \over 2T} \right) \gg 1 $.
Thus in the high temperature limit or -- turning the argument
around -- for low frequency modes, i.e.
for $k_0 \ll T $,
the classical solution yields a good approximation to the full quantum case.
To be more precise: In simulations one has to solve the classical Langevin 
equation (\ref{eq:eomcl2}) and calculate $n$-point functions by averaging over the
random sources. This has been raised in \cite{CG97}.
We would also like to refer to the works in \cite{Aa96}
for some practical examples regarding the mathematical
evaluation of expectation values in the transition from quantum
to classical field theories using dimensional reduction techniques.

One can also write down the equations of motion for the
quantum two-point functions with external noise \cite{GL98a} by introducing
the 'noisy' propagators
\begin{equation}
  \label{eq:dxi}
D_\xi^{ab}(x_1,x_2) := -i \langle \phi^a(x_1) \,\phi^b(x_2) \rangle_\xi  \, .
\end{equation}
One recovers
that $D_\xi^{\rm ret}$ and $D_\xi^{\rm av}$ obey the same equations
of motion as $D^{\rm ret}$ and $D^{\rm av}$, respectively,
and are thus the same.
Only the equation of motion for $D_\xi^<$ and hence for the occupation number is 
modified compared to the one for $D^<$. 
Averaging this equation over the noise fields according to
(\ref{eq:defavdbl}) one indeed rederives the Kadanoff-Baym equation
(\ref{eq:eomklsaI}).
This demonstrates that the Kadanoff-Baym equation can be interpreted
as an ensemble average over fluctuating fields which are subject to noise, the 
latter being correlated by the sum of self energies $\Sigma^<$ and $\Sigma^>$,
i.e.~from a transport theoretical point of view the sum of production and 
annihilation rate.
We want to note once more that the `noisy'
or fluctuating part denoted by $I$ inherent to the structure of the
Kadanoff-Baym equation (\ref{eq:kadbaym}) guarantees that the modes
or particles become correctly (thermally) populated, as can be
realized by inspecting (\ref{eq:finpnd}) or (\ref{eq:finpndweak}).

What is changed, if we replace 
our toy model of a free system coupled to an external heat bath
by a self-coupled and thus nonlinear closed system?
In an interacting field theory of a closed system the Kadanoff-Baym
equations formally have exactly the same structure as in our toy model. The
important difference, however, is that the self energy operator is now
described fully (within the appropriate approximative scheme) by the
system variables, i.e.~it is expressed as a convolution of various two-point
functions. Hence, an underlying simple stochastic process,
as in our case an external
stochastic Gaussian process, cannot really be extracted. However, we
emphasize again that the emerging structure of the Kadanoff-Baym
equations is identical. The decomposition of the
self energy operator into its three physical parts (mass shift $s$, damping $a$, 
and fluctuation kernel $I$) can immediately be
taken over. Hence these three parts keep their clear physical meaning
also for a nonlinear closed system.

We close our discussion by noting that one can also pursue to
derive a standard kinetic transport equation for the (semi-classical) phase-space distribution
$f(\vec{x}, \vec{k},t)$ including fluctuations \cite{GL98a}.
In analogy to (\ref{eq:boltz}) one gets
\begin{eqnarray}
  \label{eq:boltzxi}
\lefteqn{k_\mu \partial^\mu_X f_\xi (\vec{x},\vec{k},t)
= 
{1\over 2}  \left(
i\bar \Sigma^<(X,k) \, [f_\xi(\vec{x},\vec{k},t)+1]  \right. }
\nonumber \\ && \left. \left.
- i\bar\Sigma^>(X,k) \,
f_\xi(\vec{x},\vec{k},t)  \right)
\, \right|_{k_0=\omega^0_{\vec{k}}}  
+ \omega^0_{\vec{k}} \, {\cal F}_{\xi }(\vec{x}, \vec{k},t) 
\,.  
\end{eqnarray}
This derived kinetic transport process
(\ref{eq:boltzxi})
has the structure of the phenomenologically inspired
{\em Boltzmann-Langevin equation} \cite{Zw69}:
In order to describe
fluctuations around the average, Bixon and Zwanzig \cite{Zw69} postulated
a stochastic (classical) Boltzmann equation
in analogy to the Langevin equation for a Brownian particle.
By adding a fluctuating
collision term to the linearized Boltzmann equation (of the form of equation
(\ref{eq:boltz})) they obtained the following scheme
\begin{equation}
  \label{eq:spacehomxi1}
\frac{d}{dt} f_\xi(t) =
- \bar\Gamma (f_\xi (t)-f_{\rm eq } )  +  {\cal F}^{\rm phen}_{\xi }(t)
\end{equation}
for a system near equilibrium.
The correlation function of the fluctuating collision term
was guessed on the basis of the fluctuation-dissipation theorem.
Also it was shown that the stochastic Boltzmann equation provides a basis
for describing hydrodynamic fluctuations. On the other hand, the equations
of Bixon and Zwanzig were not derived but obtained
on the basis of intuitive arguments.
Instead, our approach carried out in \cite{GL98a} has to be considered
as a clear derivation from first principles. Indeed it shows (nearly)
a one to one correspondence to the phenomenologically introduced scheme.
However, also some severe interpretational difficulties in the interpretation
of the fluctuating phase-space density remain. We refer the interested reader
to our discussion in \cite{GL98a}.

\section{Nonperturbative resolution of pinch singularities} \label{sec:pinch}

We have seen in the previous sections that modes or quasi-particles
become thermally populated by a {\em non-perturbative} interplay
between noise and dissipative terms entering the Kadanoff-Baym equations.
It is non-perturbative as the equations of motion explicitly resum
the self energy contributions by a Schwinger-Dyson equation defined on
the real time path contour.
In addition we have observed in 
eq.~(\ref{eq:finpndweak})
that seemingly ill-defined expressions of the form $'0'/'0'$ are well-behaved
in the sense of a weak coupling limit.

Rather similar ill-defined expressions
result from so called pinch singularities within the context of strictly
perturbative expansions in
real time non-equilibrium field theory first raised in ref. \cite{AS94}.
It is the purpose of this
section to demonstrate explicitly how pinch singularities are absent
within a necessary non-perturbative context \cite{GL98a,GL98b}.

\begin{figure}
\centerline{\epsfxsize=8cm \epsfbox{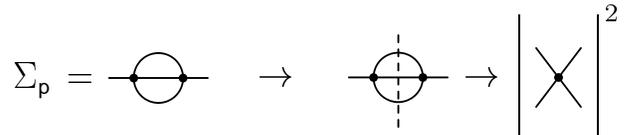}}

\caption{Lowest order self energy term in $\phi^4$-theory 
which contributes to the pinch problem (sunset diagram). 
The imaginary part of the sunset diagram can be identified with
a scattering amplitude.}
\label{fig:sunset}
\end{figure}

Employing the diagrammatic CTP rules potential `pinch singularities'
can arise in {\em strictly perturbative} expressions.
Consider in the following a weakly interacting scalar
$\phi^4$-theory.
The initial state in the far past
(assuming a homogeneous and stationary system)
is prepared by specifying the momentum occupation number $\tilde{n}(\vec{p})$ of the
(initially non interacting) onshell particles. Note that this occupation number
depends only on the three momentum $\vec{p}$.
The occupation number $\tilde{n}(\vec{p})$ enters
the (free) propagator
\begin{eqnarray}
D_0^< (p) & = &  - 2\pi i
\, {\rm sgn}(p_0) \,\delta (p^2-m^2)
\nonumber \\ && \times
\left[
\Theta(p_0) \tilde{n} (\vec{p})
-  \Theta(-p_0) (1+\tilde{n} (\vec{p})) \right]
\label{D0<noneq}
\end{eqnarray}
The free retarded and advanced propagator are given by
\begin{eqnarray}
D_0^{\rm ret/av}(p) &=& \frac{1}{p^2-m^2 \pm i \epsilon \, {\rm sgn}(p_0)} \, .
\label{D0ret}
\end{eqnarray}
A specified self energy insertion $\Sigma _{\rm p}$ in a {\em strictly perturbative}
expansion is given by a convolution of these free propagators.
As a characteristic example the
`sunset' graph arising in scalar $\phi ^4$-theory is
illustrated in fig. \ref{fig:sunset}.
We choose this particular graph
as an example since the self energies $\Sigma _{\rm p} ^{</>} (\vec{p},p_0=E_p)$ do not
vanish onshell.
The strictly perturbative corrections to the propagators
within the context of the Schwinger-Keldysh formalism take the form
\begin{eqnarray}
D^{\rm ret/av} & = & D_0^{\rm ret/av}  + 
D_0^{\rm ret/av}\Sigma _{\rm p} ^{\rm ret/av} D_0 ^{\rm ret/av}  + \ldots \nonumber \\
& =: & D_0^{\rm ret/av}  +  \Delta D^{\rm ret/av}  \,,
\label{DSDret1} \\
D^{<} & = & D_0^{<}  + 
D_0^{\rm ret}\Sigma _{\rm p}^{\rm ret}D_{0}^{<}  +
D_0^{\rm ret}\Sigma _{\rm p}^{<}D_0^{\rm av}
\nonumber \\ && 
{}+ D_{0}^{<} \Sigma _{\rm p}^{\rm av}D_0^{\rm av}  + \ldots
=: D_0^{<}  +  \Delta D^{<} \, .
\label{DSD<1}
\end{eqnarray}
As $D_0^{\rm ret}$ contains a pole
at $p_0 = \pm E_p -i\epsilon $ and $D_0^{\rm av}$ a pole
at $p_0 = \pm E_p +i\epsilon $ the product of both in the expression
for $\Delta D^< $ is ill-defined,
if $\Sigma _{\rm p}^< (\vec{ p}, p_0 = E_p=\sqrt{m^2+\vec{p}^2})$ does {\em not}
vanish onshell. Transforming such an expression back into a time representation,
the contour has to pass between this pair of two infinitely close poles.
In fact, all three contributions to $\Delta D^{<} $ are ill-defined.
On the other hand the perturbative corrections $\Delta D^{\rm ret/av}$
to the free retarded/advanced propagator are free of any pinch singularities
as the emerging poles are all located at the same side of the contour. 
The three perturbative corrections can be further rearranged \cite{AS94,GL98a}
to extract the part carrying the pinch singularity
\begin{eqnarray}
\label{DSD<2} 
\lefteqn{\Delta D^{<}_{\rm pinch} (p)  = } \\ &&
D_0^{\rm ret}(p)
\left[
\Theta (p_0) \left(
(1+\tilde{n} (\vec{p})) \Sigma _{\rm p} ^<(p) - \tilde{n} (\vec{p}) \Sigma _{\rm p} ^>(p)
\right)  \right.   \nonumber \\ 
&& \phantom{mmm} \left.  +  \Theta (-p_0) \left(
(1+\tilde{n} (\vec{p})) \Sigma _{\rm p} ^>(p) - \tilde{n} (\vec{p}) \Sigma _{\rm p} ^<(p)
\right)
\right]
D_0^{\rm av} (p) \, .  \nonumber 
\end{eqnarray}
The expression in the square brackets
is familiar from standard kinetic theory (compare with \ref{eq:boltz}).
Apart from a trivial factor one can interpret
\begin{equation}
\label{netrate}
\Gamma_{\rm eff} (\vec{p})  := 
\frac{1}{2 E_p}
\left. \left[
(1+\tilde{n} (\vec{p})) i \Sigma _{\rm p} ^>(p) - \tilde{n} (\vec{p}) i \Sigma _{\rm p} ^<(p)
\right]
\, \vphantom{\int} \right\vert_{p_0=E_p}
\end{equation}
as the net effective rate for the change of the occupation number per time.
For an equilibrium situation the occupation number
is given by the Bose distribution and the self energy insertions fulfill
the KMS condition (\ref{eq:heat}). Hence, for the equilibrium case
the whole bracket exactly vanishes and no pinch singularities emerge
within a strictly perturbative expansion.
It was observed and proven already by Landsman and van Weert that such ill-defined
terms do cancel each other, to each order in perturbation theory, if the
system stays at {\em thermal} equilibrium \cite{La87}.
This is not the case for a general non-equilibrium configuration
This seemingly `severe' problem arising for systems out of equilibrium was first
raised by Altherr and Seibert \cite{AS94}.

In a subsequent paper Altherr
\cite{A95} tried to `cure' this problem by hand by introducing a finite width for
the `unperturbed' free CTP propagator $D_0$ so that the expressions
are at least defined in a mathematical sense. Such a procedure, of course,
already represents some mixing of non-perturbative effects within the
`free' propagator.
As we have seen from our discussions
in the previous sections a damping term (resulting in a non vanishing width)
and the associated noise guarantee that the propagator becomes thermally
populated at long times. Hence, for any system which moves towards thermal
equilibrium and thus behaves dissipatively, the full propagator
must have some finite width (due to collisions or more generally due to
damping).
Within his modified perturbative approach,
Altherr \cite{A95} also showed
that seemingly higher order diagrams do contribute to a lower order in the
coupling constant, as some of the higher order diagrams
involving pinch terms will receive factors of the form
$1/\Gamma ^n, \, n\geq 1$ reducing substantially the overall power in the
coupling constant.
In his particular case Altherr investigated the dynamically generated
effective mass (the `tadpole' contribution) within standard $\phi^4-$theory.
(For the hard modes the onshell damping $\Gamma $ is of the order of $o(g^4 T)$.)
Therefore he concluded that power counting arguments might in fact
be much less trivial for systems out of equilibrium.
We will come back to his observation below.

Comparing (\ref{DSD<2}) and (\ref{netrate}) with the Boltzmann equation
(\ref{eq:boltz}) one is already tempted to speculate that the occuring
pinch singularity must be connected to the change in time in occupation number.
Hence, we will first now elaborate on the
{\em physical reason} for the occurrence of pinch singularities.
Within standard scattering theory one would think about
the probability for a particle of some initial momentum state to be
scattered into another momentum state. Therefore we ask, 
how the occupation number $\tilde{n}$ has changed after a long time. 
The occupation number for the out-states can be readily extracted
from $D^{<}$ by means of the formula \cite{GL98b}
\begin{eqnarray}
\lefteqn{n(\vec{p},t\rightarrow \infty )^{\rm (out)} = 
\langle a^{\dagger\, {\rm (out)} }_{\vec{p}} a_{\vec{p}}^{\rm (out)} \rangle  }
\nonumber \\
& = &
\left( \frac{E_p}{2} +
\frac{1}{2E_p} \frac{\partial }{\partial t} \frac{\partial }{\partial t'}
+ \frac{i}{2}  ( \frac{\partial }{\partial t} -\frac{\partial }{\partial t'} )
\right) \frac{1}{V}
\nonumber \\ && \left. 
\times
\int d^3x \int d^3y \, e^{i\vec{p} \vec{x}} e^{-i\vec{p} \vec{y}}
\left( i D^<(\vec{y},t;\vec{x},t') \right) \, \right| _{t'=t}  \, .
\label{partnumb1}
\end{eqnarray}
We now assume that the interaction is switched on at a time
$t=-T/2$ and switched off at $t=T/2$, i.e.~we replace $\Sigma _{\rm p}^{</>}(x_1,x_2)$ by
$$
\Theta({\textstyle {T \over 2}} -t_1) \, \Theta({\textstyle \frac{T}{2}}-t_2) \,
\Sigma _{\rm p}^{</>}(x_1,x_2) \,
\Theta(t_1+{\textstyle \frac{T}{2}}) \, \Theta(t_2+{\textstyle \frac{T}{2}})
$$
and assume that the duration time $T$ is large but finite.
This procedure regulates the pinch singularities 
to a finite value. Invoking a couple of manipulations \cite{GL98b}
one ends with the rather `obvious' result
\begin{eqnarray}
\Delta n(\vec{p})^{\rm (out)}_{\rm pinch}
& \approx & \Gamma _{\rm eff} (\vec{p}) 
\int \! \frac{dp_0}{2\pi } \,
\frac{4}{(p_0 -E_p)^2} \, \sin ^2 \! \! \left(\frac{T}{2}(p_0 -E_p) \right)
\nonumber \\
\label{delpartb}
& \approx &
\Gamma _{\rm eff} (\vec{p}) \cdot T  \, .
\end{eqnarray}
We thus have demonstrated the bridge
between the occurrence of pinch singularities within the context of the CTP
formalism and {\em Fermi's golden rule} in elementary quantum
scattering theory. The effective rate $\Gamma _{\rm eff}$ is therefore analogous
to the {\em transition probability per unit time}.
Indeed one can easily understand 
in physical terms that one has to expect such a singularity in perturbation theory:
Staying strictly within the first order contribution the particles remain
populated with the initially prepared non-equilibrium occupation number (since this
quantity enters the {\em free} propagator (\ref{D0<noneq})) and
scatter for an infinitely long time. Therefore,
the resulting shift $ \Delta n(\vec{p})^{\rm (out)}$
should scale with $\Gamma _{\rm eff} (\vec{p}) \cdot T $ with
$\Gamma _{\rm eff} (\vec{p})$ held fixed.
However, looking at the
Boltzmann equation (\ref{eq:boltz}),
the occupation number and the collision rate do not stay constant during the 
dynamical evolution of the system, but will be changed on a timescale
of roughly $1/\Gamma $. The quasi-particles are not really assymptotic states.

Subsequently, pinch singularities are formally cured by a resummation procedure.
The propagators will become dressed and supplemented
by a finite (collisional or more generally damping) width.
This represents already a non-perturbative effect which only
can be achieved by a resummation of Dyson-Schwinger type.
As a first attempt (proposed by Baier et al.~\cite{Ba97}), 
one might resum the full series of (\ref{DSDret1},\ref{DSD<1})
using the self energy $\Sigma _{\rm p}$.
With the definitions
$\Gamma _{\rm p}(\vec{ p}, p_0) :=
\frac{i}{2 p_0 }[\Sigma _{\rm p}^>(\vec{ p},p_0 ) - \Sigma _{\rm p}^<(\vec{ p},p_0 ) ]$ and
${\rm Re}\Sigma _{\rm p} := {\rm Re}\Sigma _{\rm p}^{\rm ret}  = {\rm Re}\Sigma _{\rm p}^{\rm av} $
one finds \cite{GL98a}
\begin{eqnarray}
\label{D<}
D^{<} & = &
D^{\rm ret} \left( (D_0^{\rm ret})^{(-1)} D_0^< (D_0^{\rm av})^{(-1)} \right) 
D^{\rm av}
\, + \,
D^{\rm ret} \Sigma _{\rm p}^< D^{\rm av}
\nonumber \\ &= &
(-2i)
\frac{p_0 \Gamma _{\rm p}}{(p^2-m^2-{\rm Re} \Sigma _{\rm p} )^2 +
p_0^2 \Gamma _{\rm p} ^2 }
\, \underbrace{\frac{\Sigma _{\rm p}^< }{\Sigma _{\rm p}^> - \Sigma _{\rm p} ^<}}_{=:n_{\Sigma }
(\vec{ p}, p_0 )}   \, .
\end{eqnarray}
(The first term on the r.h.s. of the first equation represents a boundary term
\cite{GL98a,DuB67} of the (initial) propagator $D_0^<$.
If there would be no interaction, i.e. $\Sigma _{\rm p} \equiv 0$, then
this boundary term just gives the free and undisturbed propagator.
On the other hand, if $\Sigma _{\rm p}$ is {\em nonvanishing} onshell,
as assumed, this boundary
term vanishes within the carried out resummation \cite{GL98a}.
Only the second term then contributes and yields the result
stated in the second line of (\ref{D<}).)

Hence the resummation of the series (\ref{DSD<1})
of ill-defined terms
results in a well-defined expression.
The quantity $n_{\Sigma } (\vec{ p}, p_0 )$
has to be interpreted as the `occupation number' demanded
by the self energy parts \cite{GL98a}.
If the equilibrium KMS conditions (\ref{eq:heat}) apply for the self energy part,
then $ n_{\Sigma } (\vec{ p}, p_0 )$
$
{\longrightarrow }$ $n_B(p_0 ) $
becomes just the Bose distribution function.
For a general non-equilibrium situation, however, this factor
deviates from the Bose distribution.
Comparing (\ref{D0<noneq}) with (\ref{D<}),
the astonishing thing to note is that in fact
the (initial) non-equilibrium distribution $\tilde{n}$
has dropped out completely!

Calculating $\Sigma _{\rm p}$ on
a purely perturbative level the initial occupation number $\tilde n$ enters via
the free propagator (\ref{D0<noneq}). This however cannot be the whole truth
in a dynamically evolving system.
It is important to make sure that such a 
system is prepared at some {\em finite} initial time $t_0$.
(If $t_0$ would be taken as $t_0 \rightarrow - \infty$ the system would already
have reached equilibrium long time ago.
Time reversal symmetry is
explicitly broken, so that the propagators in principle
have to depend on
both time arguments explicitly before the system has reached a final
equilibrium configuration. Therefore the simple use of
Fourier transforms, as carried out in \cite{AS94,A95,Ba97}, and which in fact
has led to the pinch singularities in (\ref{DSD<2})), is dubious \cite{GL98a}.)
The initial out of equilibrium distribution
$\tilde{n}(t_0)$ cannot stay constant during the evolution of the system
as it has to evolve towards the Bose distribution.
As long as the system is not in equilibrium,
the propagator thus cannot be stationary.
In addition, the self energy parts
$\Sigma^<$ and $\Sigma^>$ do also evolve with time. Hence they should depend
on the evolving distribution function and not persistently on the
initial one, $\tilde{n}$, which enters $\Sigma _{\rm p}$ in (\ref{D<}).
Thus, the resummation of (\ref{D<}) does not cover all relevant contributions.
The self energy operators must
be evaluated using the fully dressed and temporally evolving
one-particle propagators.

\begin{figure}
\centerline{\epsfxsize=8cm \epsfbox{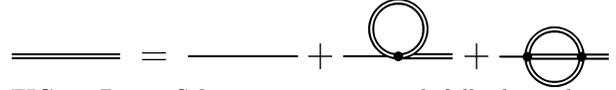}}

\caption{Dyson-Schwinger equation with fully dressed propagators (skeleton 
expansion).}
\label{fig:full}
\end{figure}

The solution to these demands is, of course, the description
of the system by means of appropriate (quantum) transport equations
like the Kadanoff-Baym equations.
Graphically this is illustrated in fig.~\ref{fig:full}.
In addition to the sunset diagram we have
also included the Hartree diagram there
which in a perturbative scheme is the one which arises first. 
The difference to the resummation of (\ref{D<}) is the fact
that the propagators entering into the self-energy operators are now
also the fully dressed ones.

Unfortunately, the full quantum transport
equations are generally hard to solve and thus are not so much of practical use.
Yet one need not be that pessimistic. For weak coupling,
i.e.~if the damping width is sufficiently small
compared to the quasi-particle energy one can take the Markov
approximation to obtain standard kinetic equations.
For the situation of fig.~\ref{fig:full}
one gets 
\begin{eqnarray}
\lefteqn{\left( E_p \partial _t -\vec{p}\partial_{\vec{x}}
- \partial_{\vec{x}} m(\vec{x},t) \partial_{\vec{p}} \right)
f(\vec{x},t;\vec{p}) = }
 \\
&&  {1\over 2} \left[
i \Sigma^<(\vec{x},t;\vec{p},E_p) \, (f(\vec{x},\vec{p},t)+1)
- i \Sigma^>(\vec{x},t;\vec{p},E_p) \, f(\vec{x},\vec{p},t)
\right] \nonumber
\end{eqnarray}
Here $f$ denotes the semi-classical non-equilibrium phase-space distribution
of quasi-particles.
$m(\vec{x},t)$ denotes the sum of the bare and the dynamical
(space time dependent) mass generated by the Hartree term.
Within this Markovian approximation the fully dressed and resummed
propagators are given by
\cite{GL98a}
\begin{eqnarray}
\lefteqn{D^{\rm ret/av}(\vec{x},t;p)  \approx }  \nonumber \\ 
&& \frac{1}{p^2 - m^2(\vec{x},t) - {\rm Re}\Sigma (\vec{x},t;p)
\pm ip_0 \Gamma (\vec{x},t;p)} \, ,
\label{Dfullret}
\\
\lefteqn{D^{<}(\vec{x},t;p)  \approx 
(-2i) \, f(\vec{x},t;\vec{p}) }
\nonumber \\ && \times
\frac{p_0 \Gamma (\vec{x},t;p)}{(p^2-m^2(\vec{x},t) -
{\rm Re} \Sigma (\vec{x},t;p) )^2 + p_0^2 \Gamma ^2(\vec{x},t;p)  }
\, .
\label{Dfull<}
\end{eqnarray}
We emphasize that
in (\ref{Dfull<}) the {\em instantaneous}
non-equilibrium phase space distribution function
$f(t)$ enters and not the initial one, $\tilde n$. 
The dynamically generated mass as well as the
collisional self energy contribution $\Sigma $ can thus be evaluated
with these propagators.
Higher order terms leading to the pinch singularities
are explicitly resummed and lead to finite and transparent results.

One can now understand the observations made by Altherr \cite{A95}.
He has found, starting from some non-equilibrium distribution $\tilde{n}$,
that higher order diagrams contribute to the same order
in the coupling constant as the lowest order one. Indeed, in his investigation,
the particular higher order diagrams where
nothing but the perturbative contributions
of the series in (\ref{DSD<1}) for the dressed
or resummed one-particle propagator $D^<$.
The only difference is that he has employed a `free' propagator
modified by some finite width in order that each of the terms in the series
(\ref{DSD<1}) becomes well defined.
The reason
for the higher order diagrams to contribute to the same order
is that the initial out-of-equilibrium distribution
$\tilde{n}$ cannot stay constant during the evolution of the system as it has
to evolve towards the Bose distribution.
If $\tilde{n} - n_B$ is of order $o(1)$,
it is obvious that there must exist contributions which perturbatively attribute 
to the
temporal change of the distribution function and contribute to the same order
$o(1)$.
In fact, in our prescription (\ref{Dfull<}),
$\tilde{n}$ has simply be substituted by the actual phase space distribution
$f$. Then calculating e.g. the tadpole diagram,
as discussed in the particular case of \cite{A95},
one has to stay within lowest order in the skeleton expansion,
but with the fully dressed propagator.

\section{Summary and conclusions} \label{sec:summary}

Our study provides new intuitive insight into
non-equilibrium quantum field theory and the process of thermalization.
In our discussions we have elucidated
on the stochastic aspects inherent to the (non-) equilibrium quantum transport
equations, the so called Kadanoff-Baym equations.
For this we have started
with a simple model, where we have coupled a free scalar boson quantum field
to an external heat bath with some given temperature $T$.
We have isolated a
term denoted by {\em I} which solely characterizes the
(thermal and quantum) fluctuations inherent to the underlying transport process.
By introducing a stochastic
generating functional the emerging stochastic equations of motion
can then be seen as generalized (quantum) Langevin processes.
Long wavelength modes with momenta and energies $|\vec{k}|$, $\omega$ $\ll T$
then behave
as classical propagating modes for a weakly interacting theory.
The important observation is that there 
the average occupation number of the
soft modes becomes large and approaches the classical equipartition limit.
We hope that our detailed analysis of the real time description
of the soft modes within the Schwinger-Keldysh formalism can attribute
to this subject. The understanding of the behavior of the soft modes is crucial
e.g.~for the issue of 
baryon number violation 
in the hot electroweak theory \cite{HS97}.

We also have demonstrated
how so called pinch singularities are regulated within the
non-perturbative context of the thermalization process. 
These singularities do (and have to) appear in the
perturbative evaluation of higher order diagrams within the CTP description
of non-equilibrium quantum field theory.
Their occurrence signals the occurrence of (onshell) damping or
dissipation. This necessitates in the description of the evolution of
the system by means of non-perturbative transport equations.
In the weak coupling regime
this corresponds to standard kinetic theory. In this case we have given a 
prescription of how the dressed propagators can be approximated in a very 
transparent form.


\end{document}